# Bent Electronic Band Structure Induced by Ferroelectric Polarization

Norihiro Oshime[1*], Jun Kano[1,2†], Eiji Ikenaga[3], Shintaro Yasui[4], Yosuke Hamasaki[4], Sou Yasuhara[4], Satoshi Hinokuma[2,5], Naoshi Ikeda[1], Mitsuru Itoh[4], Takayoshi Yokoya[6], Tatsuo Fujii[1], and Akira Yasui[3]

*[1]Graduate School of Natural Science and Technology, Okayama University, Okayama 700-8530, Japan. [2]Japan Science and Technology Agency, PRESTO, Kawaguchi, Saitama 332-0012, Japan. [3]Japan Synchrotron Radiation Research Institute, JASRI, Sayo, Hyogo 679-5198, Japan. [4]Laboratory for Materials and Structures, Tokyo Institute of Technology, Yokohama 226-8503, Japan. [5]Department of Applied Chemistry and Biochemistry, Graduate School of Science and Technology, Kumamoto University, Kumamoto 860-8555, Japan. [6]Research Institute for Interdisciplinary Science, Okayama University, Okayama 700-8530, Japan.*

Corresponding author.

[*]sc421235@s.okayama-u.ac.jp

[†]jun@psun.phys.okayama-u.ac.jp

**Abstract**

Bent band structures have been empirically described in ferroelectric materials to explain the functioning of recently developed ferroelectric tunneling junction and photovoltaic devices. This report presents experimental evidence for ferroelectric band bending, which was observed in the depth profiles of atomic orbitals of angle-resolved hard x-ray photoemission spectra of ferroelectric $BaTiO_3$ thin films. The ferroelectric bent band structure is separated into three depth regions; the shallowest and deepest regions are slightly modulated by the screening effect at surface and interface, respectively, and the intermediate region exhibits the pure ferroelectric effect. In the pure ferroelectric bent band structure, we found that the binding energy of outer shell electrons shows a larger shift than that of inner shell electrons, and that the difference in energy shift is correlated with the atomic configuration of the soft phonon mode. These findings could lead to a simple understanding of the origin of electric polarization.

Electric polarization of ferroelectric materials originates in the relative ionic displacement of a transition metal and oxygen, with inversion symmetry breaking [1]. The electric field generated by the electric polarization (polarization field) [2] causes an electrostatic potential gradient along the polarization direction in ferroelectric materials [3], causing the formation of a bent band structure. Such a graduated potential influence on the energy levels of atomic orbitals drives asymmetric electron transfer in ferroelectric tunneling junctions (FTJs) [3-8] and photovoltaic (PV) devices [9-12]. FTJs constructed of a ferroelectric thin film sandwiched by two different metals, i.e. metal (M)-ferroelectric (FE)-M junctions, exhibit electron tunneling across the barrier with an electrostatic potential gradient [5]. Switching polarization can drive the reversal of the gradient orientation [13] producing a large tunneling electroresistance (TER) effect at M-FE-M junctions. This effect dominates the tunneling magnetoresistance effect [6]. Recently, a TER effect with a magnitude of $10^4$ has been demonstrated in M-FE-heavily doped semiconductor (hS) junctions, where the hS was Nb-doped $SrTiO_3$ (NSTO) [7]. At the FE-hS interface, the combination of a wide Schottky barrier width and the small work function of the NSTO create variable depletion and accumulation states controlled by polarization reorientation [7,8]. Since band engineering of the FE-hS interface can improve the effective electron transfer, FTJs are currently seen as a promising heterostructure for ferroelectric random access memory (FeRAM), one of the several advanced ferroelectric functional devices [4,8]. In ferroelectric PV devices, a slightly wide forbidden band (2-4 eV) and the surface band bending induced by bound charges play an important role in photoconductivity [9-12]. When photons in the ultraviolet energy range excite electronic carriers from valence to conduction bands, the electronic carriers can transfer to the crystal's surface along the electronic polarization orientation, producing selective electron accumulation at the surface of the positive domain [10]. In addition, electron mobility can be enhanced by the cationic [10] and oxygen [12] vacancies that accompany surface band bending. In photochemical reactions, ferroelectric substrates offer a significant advantage in the fabrication of nanostructures such as nano metals and organic molecules, because nanometer-sized polarization domains can be patterned in positive or negative regions by an external electric field [9,10].

Electronic structures modulated by electric polarization yield a so-called ferroelectric band bending structure. Although band bending structures can be described by the effect of electric polarization on ferroelectric materials, the common band bending phenomenon has been discussed as an interfacial effect in a pn junction, which consists of electrically non-polar semiconductors such as Si and GaAs [14]. Since the band bending structure in a pn junction is derived from the different work functions of two semiconductors, it cannot be reversed by any external field. In contrast, the ferroelectric band bending structure is made possible by reversible electrical polarization. Ferroelectricity is thus a key function for future electronic devices using a bent band structure. The bent-band structure of pn junctions has been observed by transmission electron microscopy [15]; however, the actual basis of ferroelectric band bending remains obscure. One of the crucial problems is that the actual location of the band bending is not known. For example, one researcher may consider that the band bending should occur from the top to the tail of the ferroelectric thin film in the case of FTJs, while another feels that it forms at the surface of the ferroelectric thin film in PV devices. We consider that knowing the energy shift of atomic orbitals in terms of their depth profile is an important basis for understanding the electronic structure in ferroelectrics. In the present study, angle-resolved hard x-ray photoemission spectroscopy (AR-HAXPES) was shown to permit direct observation of the band bending structure in ferroelectric materials. AR-HAXPES with synchrotron radiation provides a 20-nm-deep profile of photoelectron emission spectra. Our results show that each electronic core level shifts to a higher energy in accordance with the orientation of the electrical polarization. Thus the ferroelectric bent-band structure can be attributed to three factors, i.e. pure ferroelectric band bending, a surface effect, and an interfacial effect. In pure ferroelectric band bending, the binding energies of Ti-2$p_{3/2}$ and O-1$s$ core levels shift more than that of the Ba-5$p_{5/2}$ core level in ferroelectric $BaTiO_3$ (BTO). The difference is attributed to shielding from electron configurations and correlation with phonon oscillations, which leads to a straightforward understanding of the origin of electrical polarization.

AR-HAXPES was carried out at BL47XU in SPring-8. The detailed experimental setup of the BL47XU beamline is described in Ref. [16]. Generally, the ionization cross-section decreases with increasing photon energy [17]. The required photon energy is estimated to be 8 keV when taking into account the escape depth of each atomic orbital. In the present study, a photon energy of 7.94 keV with a bandwidth of 38 meV was obtained using the Si (111) double monochromator and the Si (444) channel cut monochromator. The X-ray beam was focused to a 30 × 40 μm$^2$ region on the sample surface. The AR-HAXPES apparatus installed in BL47XU has a wide-acceptance-angle objective lens ahead of the conventional HAXPES system (R-4000-VG-Scienta Co.) [16]. The angle between the AR-HAXPES apparatus and photon propagation is fixed at 90$^o$ in all experiments. The emission angle of photoelectrons depends on the escape depth as shown in Fig. 1. The objective lens has a wide acceptance angle of 64$^o$. Since angular resolution corresponds to depth resolution from the sample, photoemission detection by the objective lens produces a wide depth-dependent analysis with a resolution of 1.32$^o$ even with a one-shot and fixed optical system. On the other hand, a conventional AR-HAXPES without a wide-angle objective lens is often required to mechanically adjust the optical angle between the incident beam and the sample, a feature causes difficulty in accurate angular-resolution and beam-positioning within the micrometer domain on samples. The energy resolution was estimated to be about 0.27 eV by Au Fermi-edge measurement.

Ferroelectric oxides have small carrier concentrations, resulting in charge at the surface and making effective photoemission detection difficult. We used epitaxial ferroelectric oxide thin films grown on a conductive single-crystal substrate. Ferroelectric BTO and non-ferroelectric $\gamma$-$Al_2O_3$ (ALO) thin films 5 nm thick were deposited on (100) Nb 0.5wt% doped $SrTiO_3$ (NSTO) single-crystal substrates by pulsed laser deposition, using the 266 nm 4th-harmonic wave of a Nd:YAG laser. Deposition conditions of BTO and ALO were respectively 650°C and 700°C growth temperature, 20 mTorr and 1 mTorr oxygen pressure, and 1.3 and 2.9 J/cm$^2$ of laser energy. Our experiment aimed to investigate the contribution of polarization to band bending. In this case, simple interface using single domain BTO thin film with avoiding any ferroelectric and ferroelastic domain contributions is necessary to understanding. Crystal structures of the deposited films were confirmed by high-resolution x-ray diffraction (XRD, Smartlab RIGAKU) with a 2-bounce monochromator. The polarization direction of the BTO film was measured by piezoresponse force microscopy (PFM, MFP-3D Oxford instruments). Figures 2(a) and 2(b) show XRD $\theta$-$2\theta$ patterns of BTO and ALO films. Both films were epitaxially grown with a cube-on-cube relation between film and substrate; (001)BTO||(001)NSTO and (100)BTO||(100)NSTO, (001)ALO||(001)NSTO and (100)ALO||(100)NSTO, respectively. Rocking curves of both films measured at *002* BTO and *004* ALO are shown in Fig. 2(c) and full-width at half-maximum (FWHM) values were 0.137° (blue, dash-line) and 0.108° (black, solid-line). There were no secondary and no different orientation peaks in both films. The crystal mosaicities of both films were almost identical, as indicated by similar FWHM values. Figure. 2(d) shows a topographic image of BTO thin film measured together with PFM. BTO film has the very flat surface with a root-mean-square roughness of 0.2 nm. As shown in Fig. 2(e), positive 3 V writing on a 2 × 2 μm$^2$ area was performed for poling treatment directed at the film surface, then negative 3 V writing under 1 × 1 μm$^2$ for poling treatment directed at the substrate at the center of a 2 × 2 μm$^2$ area was also performed. Additionally, the opposite bias for the same writing configuration was applied to the same BTO film, as shown in Fig. 2(f). Both the as-deposited area and the negative-bias writing area showed the same PFM phase contrast, although contrariwise, a positive bias writing area showed the opposite PFM phase contrast. No in-plane contribution of ferroelectricity was found by PFM measurement because this BTO film with the very thinner thickness of 5 nm was grown on NSTO substrate with the fully compressive strain from STO. The reason is that lattice parameter of STO substrate is smaller than that of BTO. In the case of BTO, it is well known experimentally and theoretically that tetragonality is enhanced by compressive strain [18]. Therefore, BTO film has only single c-domain and their polarization direction is headed to the substrate. Note that, ALO thin film did not show any piezoresponse.

For AR-HAXPES, we checked spectra with angle-integrated HAXPES in survey setting (TOA = 88.3°) and then confirmed peak selection, with the result that Ti-2$p_{3/2}$, O-1$s$, Ba-3$d_{5/2}$ and Sr-3$s$ in BTO/NSTO and O-1$s$ in ALO/NSTO were selected. We analyzed the observed atomic orbitals with the following process: (1) background was subtracted by Shirley function; (2) subtracted spectrum was fitted by Voigt function (see gray curves in Fig. 3(b)); (3) the binding energy of the atomic orbital was estimated to be the center position of FWHM. Finally, the depth-dependence of the energy shift in the atomic orbital was determined.

No extrinsic charging effect was observed in our experimental data. Such a charge is often seen in photoemission experiments, where it degrades the accuracy of the data. In an extrinsic charge situation, the energy level of photoelectrons at the surface is often altered by the electric field of the surface space charge, expanding the distribution of kinetic energy and resulting in a spectral profile that increases in width as the depth decreases towards the surface. However, we confirmed that the Gaussian width of the Voigt function was constant with respect to the emission angle, i.e. the distribution of kinetic energy was not expanded. Thus we conclude that there was no surface charge on our samples in the present study. Electron beam irradiation by the flood gun was not used, to avoid spectral distortion.

Figure 3(a) shows AR-HAXPES spectra of Ti-2$p_{3/2}$ in BTO observed at various depths. Emission angles of 5° to 65° correspond to photoelectron emissions from the surface to a deeper region, respectively. In the spectrum of Ti-2$p_{3/2}$, the peak shifts to a higher binding energy region with increasing escape depth. As shown in Fig. 3(b), O-1$s$ in ALO splits into two peaks: one in the vicinity of 532.5 eV and another in the vicinity of 530.6 eV, which respectively originate from ALO and NSTO. With increasing sample depth, the O-1$s$ orbital derived from NSTO appears gradually.

For a quantitative discussion of the energy shift in atomic orbitals, we fitted the changes by superposition of Voigt functions. Figure 3(c) shows the depth-dependence of binding energies of Ti-2$p_{3/2}$ in BTO. In the internal layer of BTO (corresponding to emission angles 15°-45°), the binding energy of Ti-2$p_{3/2}$ increases monotonically with increasing depth. This energy shift behavior is consistent with a potential slope where the electric polarization points into the NSTO substrate (see Fig. 2(d) and 2(e)). Thus we can conclude that this slope appeared in the internal layer as ferroelectric band bending (FeBB) induced by electric polarization. We consider other bending contributions below. At the surface (corresponding to emission angles $5^o$-$15^o$), the energy shift is suppressed. This behavior is caused by surface screening positive charge at the BTO surface so-called surface band bending (SBB). Since surface screening charges are introduced by electric polarization in a sample's surface, surface screening positive charges cover the BTO surface. The charges generate an electric field in the direction anti-parallel to the polarization, so that the FeBB is screened and suppressed by the SBB at the sample's surface. At the BTO/NSTO interface (corresponding to emission angles $45^o$-$65^o$), the Ti-2$p_{3/2}$ orbital has an inflection point. This behavior is attributed to the formation of a Schottky barrier involving carrier compensation between BTO and NSTO, resulting in the appearance of interfacial band bending (IBB). NSTO is regarded as a metal, with a large carrier concentration of $1-2\times10^{20}$ $cm^{-3}$ and a small work function. When the polarization points into NSTO, no insulating region exists at the interface [7,8]. Therefore, band alignment forms a Schottky barrier in the BTO/NSTO interface resulting in the formation of IBB. FeBB, SBB and IBB are essential in TER [3-8] and PV [9-12]. In ALO, similar SBB and IBB behavior appear without FeBB (see Fig. 3(d) and the Supplemental Material [19]).

Based on our band-bending lineup as described above, we discuss a band bending structure induced purely by a ferroelectric contribution, i.e. FeBB. Figure 4 shows the depth-dependence of the binding energies of Ti-2$p_{3/2}$, O-1$s$, Ba-3$d_{5/2}$ and Sr-3$s$ in BTO/NSTO. The energy level of all atomic orbitals has the three components of SBB, FeBB and IBB, going from the surface to the interface between BTO and NSTO. A polarization field produces a gradual change of electrostatic potential in a ferroelectric crystal. Since such a potential has a slope due to the polarization field, the binding energy increases with increasing depth in the sample, as seen in Fig. 4. Electronic carriers in BTO move to the BTO/NSTO interface along the potential slope, so that a depletion region is formed in the inner layer of BTO. The polarization field thus remains in BTO without screening, forming FeBB states. In our experiment, an electronic structure bent by the polarization field is observed as a binding energy shift of atomic orbitals.

All atomic orbitals show similar behavior, but the magnitude of the energy shift derived from FeBB is different, as shown in Fig. 4. Below, we discuss in detail the mechanism by which FeBB produces a different energy in each atomic orbital. This behavior can be attributed to shielding by the electron configuration between the effective nuclear charge and the outer electron shell. When an electron with low binding energy is located in an outer shell orbital, the electron has a large shielding constant, so that the effective nuclear charge becomes small [20]. Low binding energy means that the Coulomb force between the electron and the nuclear charge is weak. Thus the polarization field has an effect on the electron such that the energy shift increases. Conversely, when an electron with high binding energy is located in an inner shell orbital, the electron has a small shielding constant, so the effective nuclear charge does not change [20]. Since the Coulomb force between the electron and the nuclear charge is strong, the effect of the polarization field is negligible, decreasing the energy shift. The energy shifts of Ba-3$d_{5/2}$, O-1$s$ and Ti-2$p_{3/2}$ in BTO are estimated to be 0.08, 0.11 and 0.17 eV, respectively, where the binding energies of atomic orbitals are located at 779.6, 530.5 and 459.1 eV, respectively. These shift values are in good agreement with the above discussion.

Note that the binding energy shift of Ba-3$d_{5/2}$ is small, as shown in Fig. 4, even though the 3$d_{5/2}$ orbital is located in the outer shell in the electron configuration of Ba. We argue that the energy shift is correlated with phonon oscillation in the ferroelectric crystal. The polar phonon mode in BTO is assigned to the oscillation of a Ti ion and an oxygen octahedron; the so-called Slater mode [21]. Softening of the Slater mode induces a structural phase transition together with the generation of electric polarization. This means that the displacement of a Ba ion makes no contribution to the Slater mode. There are large binding energy shifts in Ti and O ions, which have a Slater mode atomic configuration, but Ba has a small shift. In addition, Ti and O ions are hybridized, while Ba displays ionic behavior [22]. Thus, the large binding energy shift of Ti and O ions triggers hybridization. In the case of $PbTiO_3$ (PTO), the soft phonon mode is assigned to the oscillation of Pb and the $TiO_6$-octahedron, i.e. the Last mode [21,23]. We infer that the electronic structure in PTO shows a large energy shift in the core level of Pb atomic orbitals even for electrons located in inner shells. Ti and O show similar tendencies. This interpretation is consistent with the hybridizations of Pb-O and Ti-O [24]. Although the volume of electric polarization has usually been evaluated by the combination of ionic displacement and covalency in ferroelectrics, the origin of electric polarization can be simply understood from the electronic band bending structure.

In summary, we have experimentally studied the electronic band bending structure in BTO/NSTO by AR-HAXPES. Our results show that the binding energy of core levels displays a shift with depth dependence, which provides the straightforward interpretation that the ferroelectric electronic structure has three band-bending components, i.e. SBB, FeBB, and IBB. In particular, FeBB has a gradual slope along the electric polarization direction. FeBB is affected by shielding from the electronic configuration and the characteristic behavior of ions with an atomic configuration of soft phonon mode. The theory of electronic structure can guide research on ferroelectrics. Our results will allow the development of novel PV and FTJs devices using ferroelectrics.

**Acknowledgements**

The authors would like to thank Prof. N. Nakajima (Hiroshima University) for comments on the manuscript. This work was supported by the SPring-8 Budding Researchers Support Program, the Program for Advancing Strategic International Networks to Accelerate the Circulation of Talented Researchers from JSPS (R2705) and JST, PRESTO. The synchrotron radiation experiment was performed at the BL47XU of SPring-8 with the approval of the Japan Synchrotron Radiation Research Institute (JASRI) (Proposal No. 2016B1673).

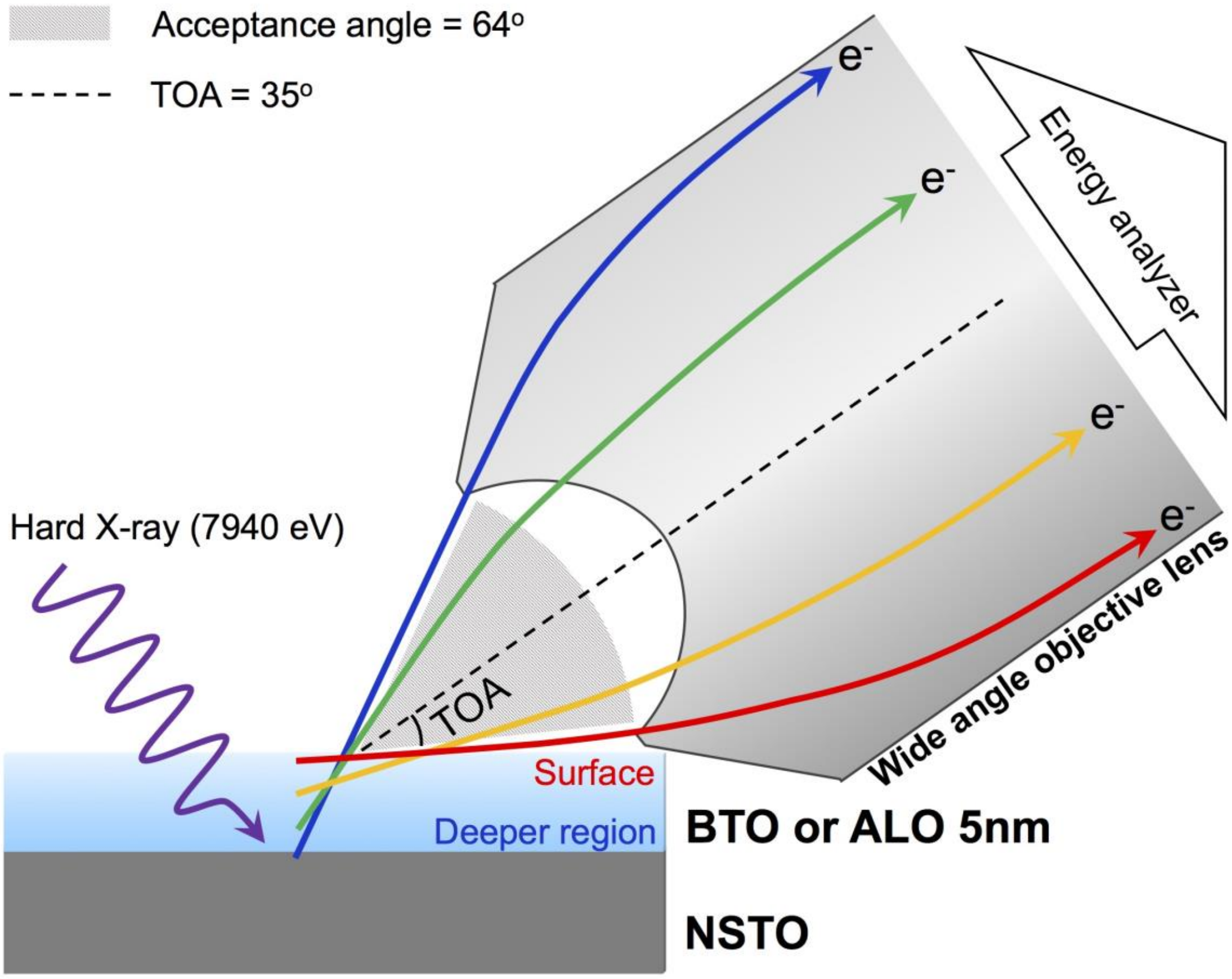

Fig. 1. Schematic picture of AR-HAXPES measurement with wide-angle objective lens for BTO (or ALO) thin film. The lens has a 64$^{o}$ acceptance angle. Take-off angle (TOA) is defined as the angle between the sample surface and the lens. The emission angle of photoelectrons increases with increasing escape depth, their TOA was determined to be 35$^{o}$.

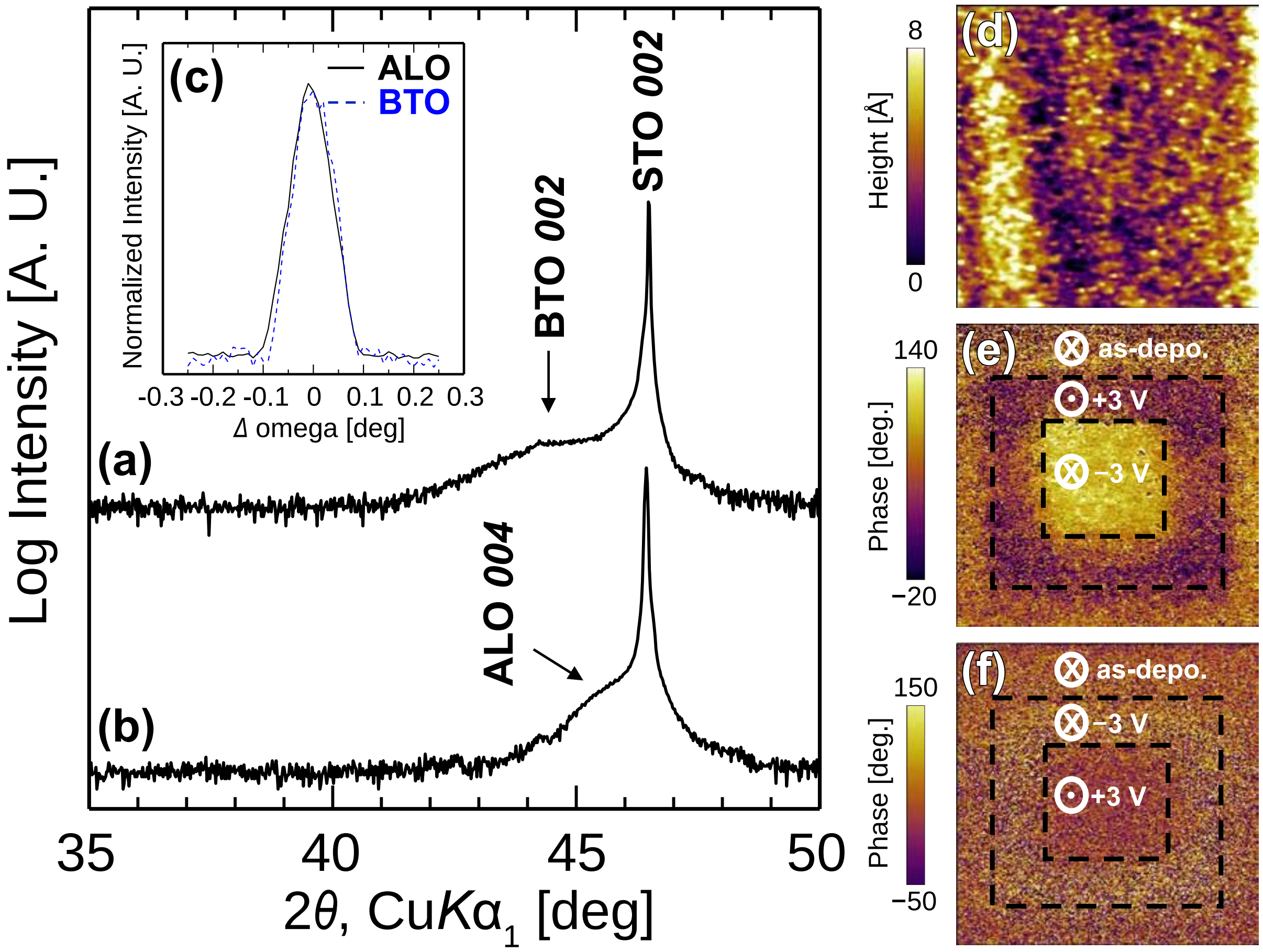

Fig. 2. X-ray diffraction patterns of (a) BTO and (b) ALO films on NSTO substrates. (c) Rocking curves measured at BTO *002* (blue, dash-line) and ALO *004* (black, solid-line) diffractions. (d) Topographic image of BTO film. Piezoresponse phase images of BTO film: (e) +3 V (2 × 2 μm$^2$ area, outside) and -3 V (1 × 1 μm$^2$ area, inside) and (f) -3 V (2 × 2 μm$^2$ area, outside) and +3 V (1 × 1 μm$^2$ area, inside) writing treatments, with a measured area of 3 × 3 μm$^2$. Light and dark regions correspond to negative and positive polarization directions, respectively.

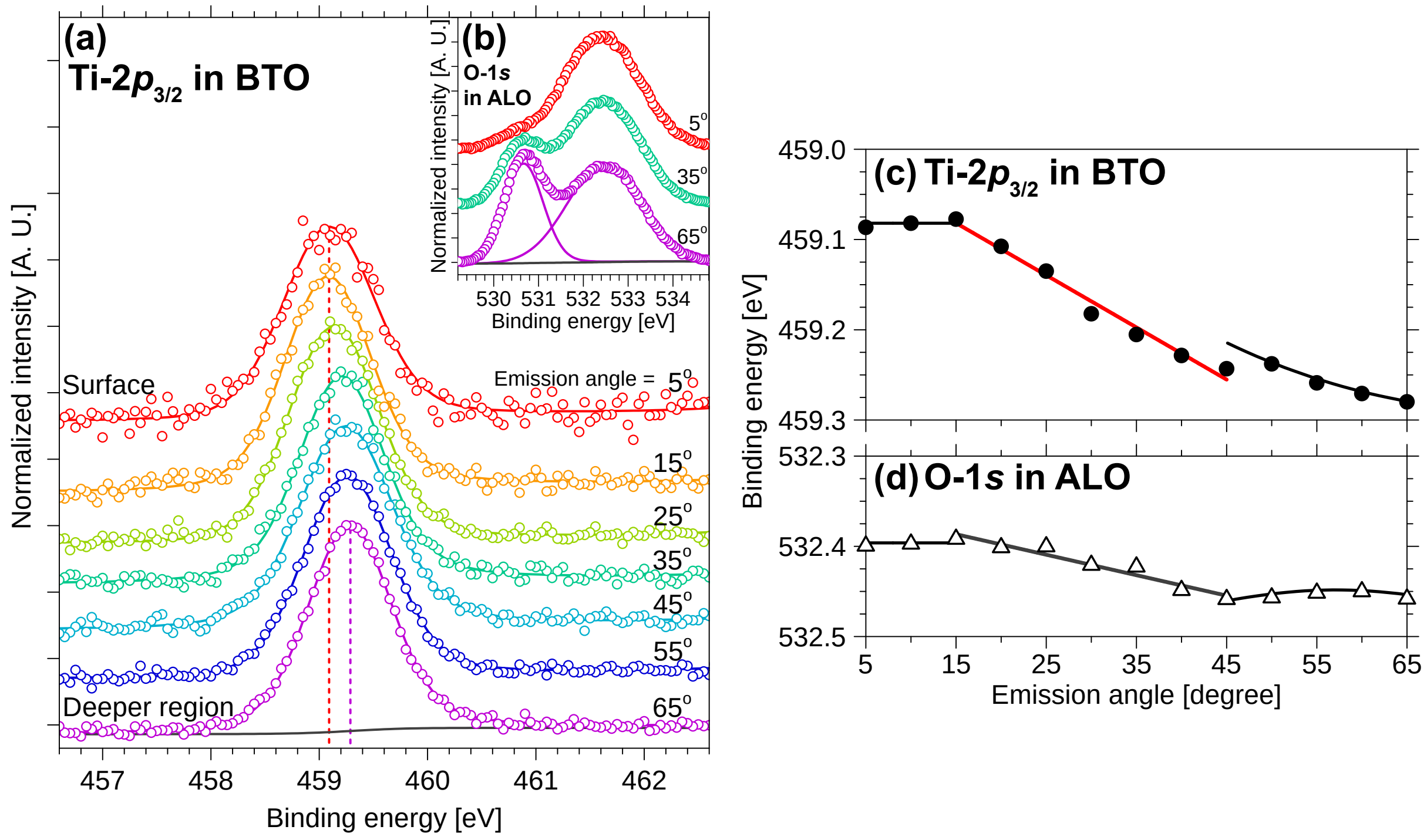

Fig. 3. AR-HAXPES spectra: (a) Ti-2$p_{3/2}$ in BTO (b) O-1$s$ in ALO. The probing depth in photoemission increases as the emission angle increases. Red and purple circles are surface and deeper regions, respectively. In the spectrum at emission angle = 65°, curves of Shirley and Voigt functions are drawn as black and purple lines, respectively. Depth dependence of binding-energy: (c) Ti-2$p_{3/2}$ in BTO (d) O-1$s$ in ALO. Black and white dots indicate the center position of FWHM of the spectrum at each emission angle. In the FeBB region of Ti-2$p_{3/2}$ in BTO, the dots were fitted by a linear approximation function, shown as a red line.

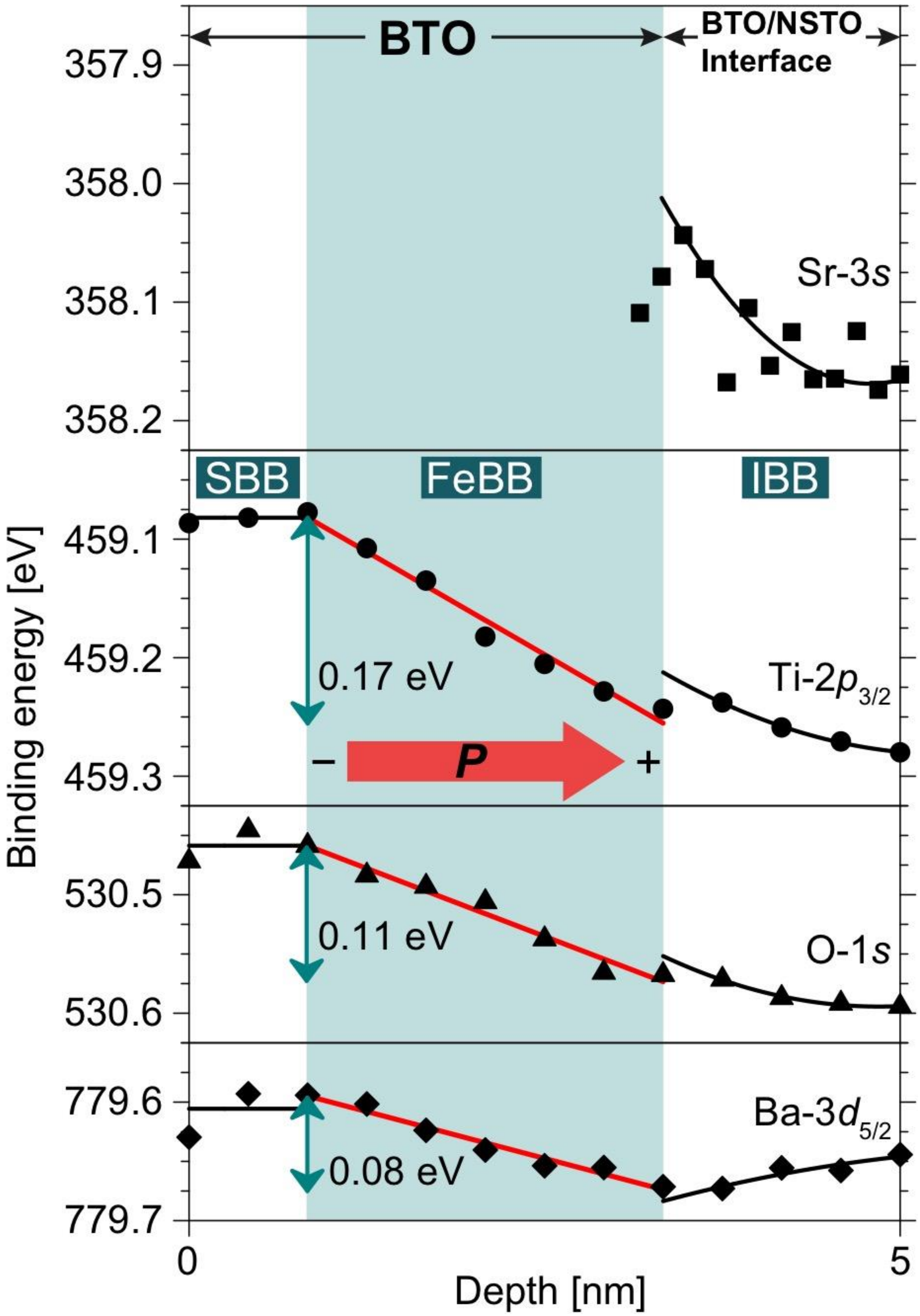

Fig. 4. The lineup of binding energies of Ti-2$p_{3/2}$, O-1$s$, Ba-3$d_{5/2}$ and Sr-3$s$ in BTO/NSTO with depth dependence. The red arrow indicates electric polarization pointing into NSTO.

# Supplementary Information: Bent Electronic Band Structure Induced by Ferroelectric Polarization

Norihiro Oshime[1], Jun Kano[1,2], Eiji Ikenaga[3], Shintaro Yasui[4], Yosuke Hamasaki[4], Sou Yasuhara[4], Satoshi Hinokuma[2,5], Naoshi Ikeda[1], Mitsuru Itoh[4], Takayoshi Yokoya[6], Tatsuo Fujii[1], and Akira Yasui[3]

*[1]Graduate School of Natural Science and Technology, Okayama University, Okayama 700-8530, Japan. [2]Japan Science and Technology Agency, PRESTO, Kawaguchi, Saitama 332-0012, Japan. [3]Japan Synchrotron Radiation Research Institute, JASRI, Sayo, Hyogo 679-5198, Japan. [4]Laboratory for Materials and Structures, Tokyo Institute of Technology, Yokohama 226-8503, Japan. [5]Department of Applied Chemistry and Biochemistry, Graduate School of Science and Technology, Kumamoto University, Kumamoto 860-8555, Japan. [6]Research Institute for Interdisciplinary Science, Okayama University, Okayama 700-8530, Japan.*

We remark a quantity of polarization field at a different thickness as below. In BTO, we obtained AR-HAXPES spectra of both 5 and 15 nm thickness samples as shown in Fig. 4 and Fig. S1, respectively. The shift of binding energy in 5 nm is larger than that of 15 nm. Generally, tetragonality of ferroelectric perovskite thin films is enhanced by compressive strain (Ref. [2]). The difference of polarized magnitude between 5 and 15 nm seems to be negligible. In this case, the slope of FeBB of 15 nm thickness is small compared with 5 nm one because the electric field is proportional to ~ $1/r^2$. This consideration consists with our experimental results.

We also confirm the difference of energy shift between ferroelectric BTO and non-polar ALO. In ferroelectric BTO (see Fig. 4), FeBB is clearly observed along to the polarization direction in the inner layer of Ti-2$p_{3/2}$ and O-1$s$ with large energy shifts. On the other hand, a shift in non-polar ALO is negligible (see Fig. S2). Actually, both of Al-2$s$ and O-1$s$ in ALO shows slight energy shift of 0.07 eV. The shift is attributed to the different charge amount between ALO surface and ALO/NSTO interface. Therefore, our experimental result sufficiently proves FeBB structure even at no demonstration of polarization switching.

## Supplemental Figures

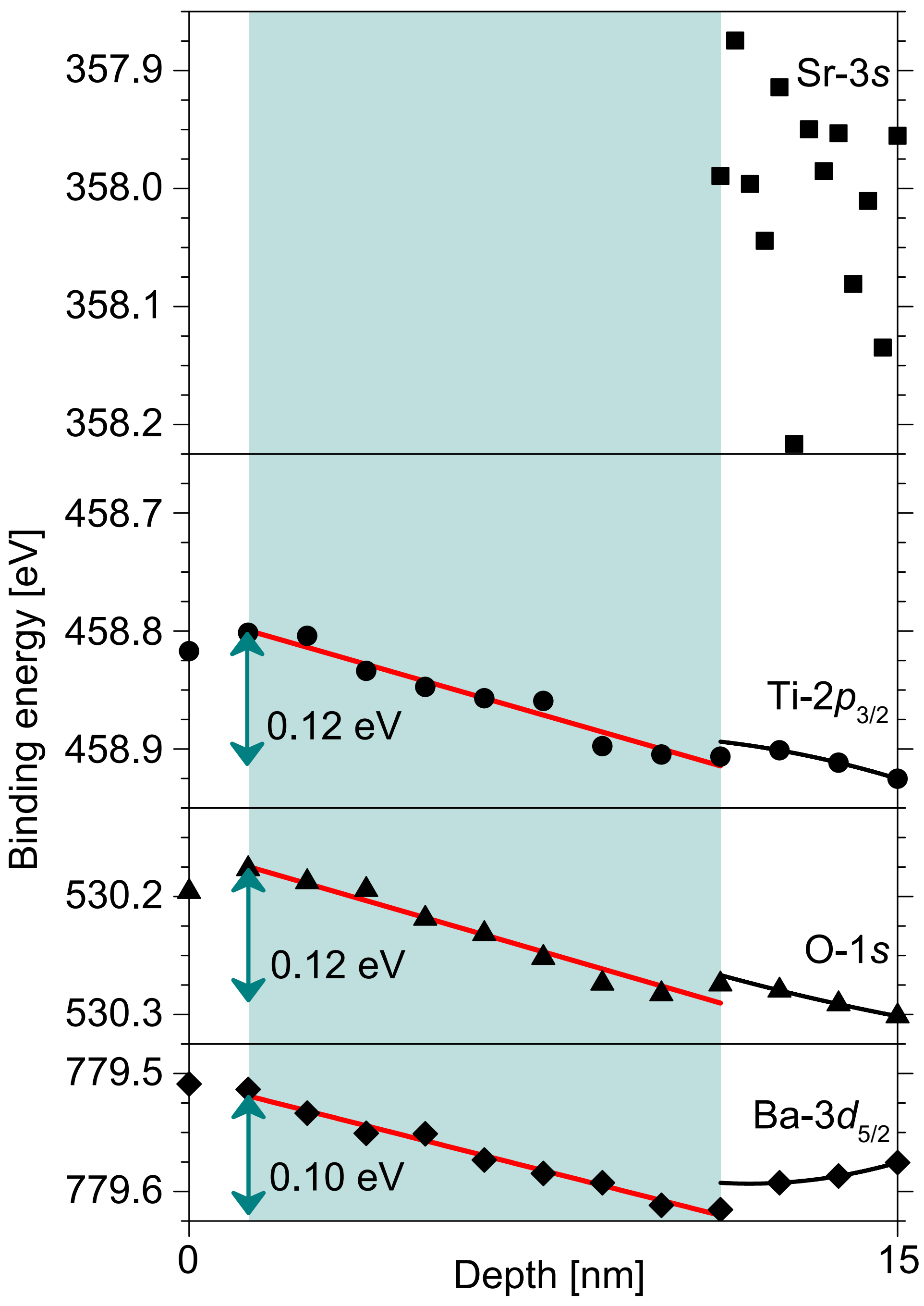

Figure S1. The lineup of binding energies of Ti-2$p_{3/2}$, O-1$s$, Ba-3$d_{5/2}$ and Sr-3$s$ in 15 nm BTO/NSTO with depth dependence.

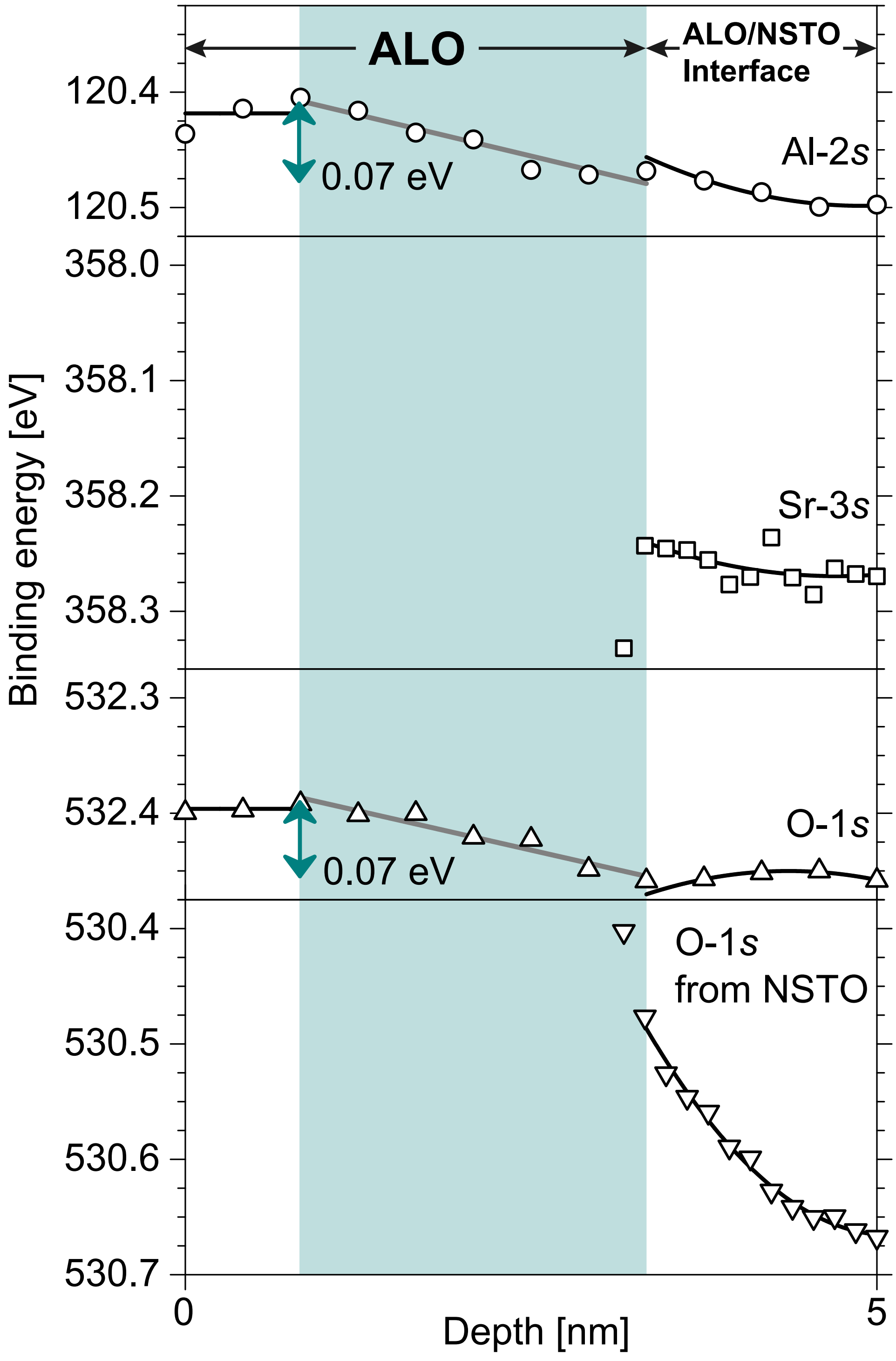

Figure S2. The lineup of binding energies of Al-2*s*, O-1*s* and Sr-3*s* in ALO/NSTO with depth dependence.